\documentclass[aps,showpacs,prd,twocolumn,nofootinbib,nobibnotes]{revtex4-2}

\usepackage{graphicx}
\usepackage{amssymb}
\usepackage{amsmath}
\usepackage{color}
\usepackage{float}
\usepackage{ulem}
\usepackage{booktabs} 
\usepackage{accents}
\usepackage{amsfonts}
\usepackage[colorlinks=true,
pdfstartview=FitV,linkcolor=blue,
citecolor=blue,urlcolor=blue,breaklinks=true]
{hyperref}
\usepackage{array}
\usepackage{float}
\usepackage[dvipsnames]{xcolor}
\usepackage{csquotes}
\usepackage{bbold}
\usepackage{units}
\usepackage{enumitem} 
\usepackage{dsfont}

\usepackage{bigints}
\usepackage{amssymb}
\usepackage{systeme}

\newcolumntype{C}[1]{>{\centering\arraybackslash}m{#1}}

\renewcommand{\eqref}[1]{\mbox{Eq.~(\ref{#1})}}

\definecolor{ForestGreen}{rgb}{0.13,0.55,0.13}

\usepackage{fixmath}

\newcommand{\orcid}[1]{\href{https://orcid.org/#1}{\includegraphics[width=10pt]{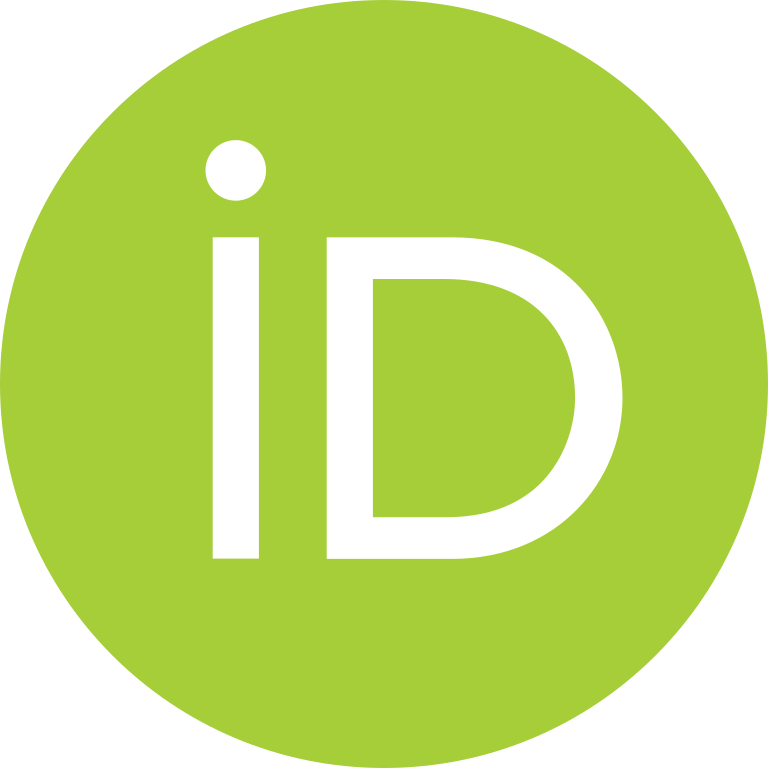}}}

\begin{document}
	
	\title{Constraining {\textit{CPT}-odd electromagnetic chiral} parameters with pulsar timing}

\author{Filipe S. Ribeiro\orcid{0000-0003-4142-4304}$^a$}
\email{filipe.ribeiro@discente.ufma.br, filipe99ribeiro@hotmail.com}
	\author{Pedro D. S. Silva\orcid{0000-0001-6215-8186}$^{a,b}$}
	\email{pedro.dss@ufma.br, pdiego.10@hotmail.com}
	\author{Manoel M. Ferreira Jr.\orcid{0000-0002-4691-8090}$^{a, c}$}
	\email{manojr.ufma@gmail.com, manoel.messias@ufma.br}
		\affiliation{$^a$Programa de P\'{o}s-gradua\c{c}\~{a}o em F\'{i}sica, Universidade Federal do Maranh\~{a}o, Campus
		Universit\'{a}rio do Bacanga, S\~{a}o Lu\'is, Maranh\~ao, 65080-805, Brazil}
		\affiliation{$^b$Coordena\c{c}\~ao do Curso de Ci\^encias Naturais - F\'isica, Universidade Federal do Maranh\~ao, Campus de Bacabal, Bacabal, Maranh\~ao, 65700-000, Brazil}
	\affiliation{$^c$Departamento de F\'{i}sica, Universidade Federal do Maranh\~{a}o, Campus
		Universit\'{a}rio do Bacanga, S\~{a}o Lu\'is, Maranh\~ao, 65080-805, Brazil}

\begin{abstract}
	The arrival time of electromagnetic signals traveling in chiral cosmic media is investigated in the context of {Maxwell-Carroll-Field-Jackiw electrodynamics}. Considering {the interstellar medium (ISM) as} a cold, ionized, chiral plasma, we derive the time delay between two traveling signals, expressed in terms of a modified dispersion measure (DM) which receives additional contribution from the chiral parameters. Faraday rotation angle is also {addressed} in this chiral plasma scenario, yielding modified rotation measures (RM). Using DMs data from five pulsars, we establish constraints on the chiral parameter magnitude at the order of $10^{-23}$ -- $10^{-22}$ GeV. On the other hand, the Faraday rotation retrieved from RM measurements implied upper constraints as tight as $10^{-36}$ GeV.

\end{abstract}
\pacs{41.20.Jb, 11.30.Cp, 03.50.Kk, 41.90.+e, 42.25.Lc}
	\maketitle

\textbf{\textit{Introduction \label{themodel1}}} -- In recent decades, radio pulsar emissions have been used to investigate a large variety of interesting topics. Electromagnetic signals, originated from pulsars and other astrophysical sources, traveling through the interstellar medium (ISM) taken as a cold plasma \cite{Lorimer-Kramer}, undergo dispersive propagation that leads to characteristic alterations in the wave propagation. For a usual cold plasma with magnetic background {${\bf{B}}_{\parallel}$ aligned to the direction of propagation}, the propagating right- and left-handed circularly polarized (RCP and LCP) transverse waves are related to the refractive indices \cite{refJACKSON, refZANGWILL}
\begin{equation}
n_{R,L}^{2}=1-{\omega_{p}^{2}}/{(\omega\left(\omega\pm\omega_{c}\right))}, \label{n_usual}
\end{equation} 
with $\omega_{p}=\sqrt{n_{e}e^{2}/m}$ and $\omega_{c}=e B_{\parallel}/m$ being the plasma and cyclotron frequencies, and $e$, $m$, $n_{e}$ the electron charge, mass and number density (given in cm$^{-3}$), respectively. {Equation (\ref{n_usual}) does not hold for the components of the magnetic field transversal to the propagation path and considers the so-called Faraday configuration, a standard approach in pulsar signal analysis that leads to magnitude estimates of the Galactic magnetic field.}

The arrival time of an electromagnetic signal traveling across a distance $d$ through ISM is defined as $t=\int_{0}^{d}(v_{g})^{-1}ds$, where $s$ is the line of sight element and $v_{g}$ the group velocity \cite{Lyne}. {For the coming light} {from pulsars, it is important to consider that the electromagnetic waves propagate through the interstellar medium (ISM) without absorption {(lossless media)}. In this sense, the ISM refractive indices need to be real, yielding real group velocities,} {which is assured when the wave frequency is large compared to the plasma frequency, $\omega\gg\omega_{p}$, with \eqref{n_usual} implying}
\begin{equation}
	v_{g}^{-1}\approx 1+\frac{\omega_{p}^{2}}{2  \omega^{2}}\pm\frac{\omega_{c}\omega_{p}^{2}}{\omega^{3}}. \label{groupvelocity}
\end{equation}
The arrival time becomes 
\begin{equation}
	t\approx d+\frac{e^{2}}{2m\omega^{2}}\text{DM}, \quad \text{DM}=\int^{d}_{0}n_{e}ds, \label{timedelay2}
\end{equation}
where the dispersion measure (DM) is defined in terms of the electron number density, $n_{e}$, assumed, in principle, not constant along the path of integration. {In general,
	the influence of the term in $\omega^{-3}$ in Eq.~(\ref{groupvelocity}) is not computed for
	the time delay, since it is smaller than the second one by the factor $\omega_{c}/\omega \sim 10^{-7}$ (for an interstellar magnetic field of the order of a few $\mu \mathrm{G}$).}  Taking the difference between the transit time of two signals (traveling at light speed $c$ and at $v_g$), the time delay, obtained from (\ref{timedelay2}), reads
\begin{equation}
	\tau = \frac{e^{2}}{2m\omega^{2}}\text{DM}, \label{timedelay3}
\end{equation}
displaying the well-known $\omega^{-2}$ behavior for electromagnetic signals\footnote{For densities much higher than the typical density of the ISM, extra terms may be included in the time delay expression, as discussed in Ref.~\cite{Kuz'min}.}. The electromagnetic time delay (\ref{timedelay3}) provides relevant information to estimate the Galactic electron distribution permeating the ISM \cite{Rybicki, Lyne}, while the DM is a key parameter for studying dispersive ISM effects along the wave path \cite{Krishnakumar-1}.

One possible way to examine the influence of Galactic magnetic fields on the pulsar signal propagation is by analyzing the Faraday rotation, a measure of birefringence. The wavenumbers associated to the indices (\ref{n_usual}), within the small density hypothesis, are
\begin{equation}
	k_{R,L}\approx \omega-\frac{\omega_{p}^{2}}{2\omega}\pm\frac
	{\omega_{c}\omega_{p}^{2}}{2\omega^{2}}.
\end{equation}
The differential phase rotation along the line of sight, 
	\begin{equation}
		\Delta\Psi=\int_{0}^{d}\left(  k_{R}-k_{L}\right)  ds=\frac{e^{3}\lambda^{2}}{4\pi^{2}m^{2}}\int_{0}^{d}n_{e}B_{\parallel} ds, \label{Faraday}
	\end{equation}
yields the polarization rotation angle $\Delta \phi$, 
	\begin{equation}
		\Delta\phi \equiv\Delta\Psi/2 =\lambda^{2}\text{RM},
	\end{equation}
	in terms of the rotation measure (RM), which is given by 
	{\begin{equation}
	\text{RM}=\frac{e^{3}}{8\pi^{2}m^{2}}\int_{0}^{d}n_{e}B_{\parallel} ds,\label{Faraday2}
	\end{equation}}{measured in $\mathrm{rad}/\mathrm{m}^{2}$.} Here, $B_{\parallel}$ is the magnetic field parallel to the line of sight (usually given in $\mu$G), while the distance $d$ is taken in pc. {Although Eq.~(\ref{Faraday2}) does not take into account transversal components of the magnetic field direction along the line of sight, it is the relevant configuration to examine the large-scale structure of the Galactic field  \cite{Lyne,Simard-Normandin}, by combining estimates of $B_{\parallel}$ for several pulsars along different lines of sight.} For instance, the RM is extensively used to estimate the magnitude and direction of the Galactic magnetic fields by examining the polarization of light coming from pulsars \cite{Han,Ng,Sullivan}.  
Radio pulsars have been observed by various telescopes, resulting in datasets for several parameters of pulsars, including DM measurements. In this context, pulsar timing is a useful technique to explore such phenomena by observing the highly regular pulses arising from pulsars \cite{Lommen}. The LOw-Frequency ARray (LOFAR) is a radio telescope consisting of an interferometric array of dipole antenna stations located in the north of the Netherlands and across Europe \cite{Haarlem}. Reference \cite{LOFAR} compiles data on DMs, flux densities, and calibrated total intensity profiles for a subset of pulsars obtained by LOFAR's high-band antennas (110-188MHz). In turn, RMs measurements have also been improved by LOFAR results for low-frequency pulsars \cite{LOFAR_RM}.

{Among possible candidates for cold dark matter, there is the QCD axion \cite{Preskill, Feng}, a pseudoscalar particle that emerges as a solution to the strong CP problem in quantum chromodynamics \cite{Peccei}. The axion coupling with the electromagnetic field is described by the term, $\mathrm{{\mathcal{L}_{axion}}}=\theta (\mathbf{E}\cdot \mathbf{B)}$, where $\theta$ represents the axion field \cite{Sekine, Tobar}. It has
been argued that axionlike particles (ALPs), which present couplings similar to the axion-photon coupling term, are leading candidates for cold dark matter \cite{Paola-ALP,Chao,Ferreira}.} In the case the axion derivative is considered as a constant vector, $\partial_{\mu}\theta=(K_{AF})_{\mu}$, the axion coupling yields the well-known Maxwell-Carroll-Field-Jackiw (MCFJ) electrodynamics,
	$\mathrm{{\mathcal{L}}}=-\frac{1}{4}G^{\mu\nu}F_{\mu\nu}    + \frac{1}{4}%
	\epsilon^{\mu\nu\alpha\beta}\left( K_{AF}\right)_{\mu}A_{\nu}F_{\alpha
		\beta} $,
where $\left(K_{AF}\right)_{\mu}$ is the 4-vector background that induces the Lorentz symmetry breaking, $F^{\mu\nu}$ and $G^{\mu\nu}$ are the electromagnetic field strengths in vacuum and in matter \cite{refPOST, Pedroo}. {The MCFJ electrodynamics \cite{CFJ,Colladay} has been investigated in a broad range of physical scenarios \cite{CFJG1,CFJG2,CFJG3}, being constrained by Schumann resonances (at the level of $10^{-20}$ GeV) \cite{Mewes-bounds-1}, laboratory-based analysis involving resonant cavities (at the level of $10^{-23}$ GeV) \cite{Yuri}, solar wind deviation data (as $10^{-24}$ GeV) \cite{Spallicci},  and by its effects on the cosmic microwave background anisotropies (at the tight limit of $10^{-44}$ GeV) \cite{Caloni}.}

{Furthermore, the MCFJ electrodynamics provides} an effective framework to describe chiral phenomena in condensed matter, such as the chiral magnetic effect (CME)\cite{Kharzeev1, Kharzeev1A, Kharzeev1B, Fukushima, Kharzeev1C, LiKharzeev} and anomalous Hall effect (AHE) \cite{ Haldane, Xiao, Huang, Liu}, often addressed in Weyl semimetals. These chiral effects are connected to the MCFJ electrodynamics, with the chiral magnetic current density being written as  $\mathbf{J}_{B}=K_{AF}^{0}\mathbf{B}$, where $K_{AF}^{0}$ plays the role of the magnetic conductivity \cite{Pedro1, PedroPRB2024A}, and the chiral vector $\mathbf{K}_{AF}$ represents the anomalous Hall conductivity in the current $\mathbf{J}_{AH}=\mathbf{K}_{AF} \times\mathbf{E}$ \cite{Qiu}. MCFJ electrodynamics has also been recently applied to address chiral cold plasmas \cite{Filipe1, Filipe2,Filipe3}, which are described by the following permittivity tensor \cite{Filipe1}:
\begin{equation}
		\tilde{\varepsilon}_{ij}=\varepsilon_{ij}\left(\omega\right) + iK_{AF}^{0} \epsilon_{ikj}k^{k}/\omega+i \epsilon_{ikj}K_{AF}^{k}/\omega,\label{ChiralPlasma}
\end{equation}
where
\begin{align}
\varepsilon_{ij}(\omega) &= S \delta_{ij} + i D \epsilon_{ij3} + (P -S) \delta_{i3} \epsilon_{j12}, 
\end{align}
{with 
	\begin{align}
	S=1 - {\omega_{p}^{2}}/{(\omega^{2}-\omega_{c}^{2})},& \quad D= { \omega_{c} \omega_{p}^{2}}/{\omega (\omega^{2} - \omega_{c}^{2})}, \\ \qquad P = &1- {\omega_{p}^{2}}/{\omega^{2}}.
	\end{align}}
In \eqref{ChiralPlasma}, the { chiral factor} $K_{AF}^{0}$ and chiral vector $\mathbf{K}_{AF}$ lead to modified effects (e.~g., birefringence and dichroism) in magnetized \cite{Filipe1,Filipe2} and unmagnetized plasma \cite{Filipe3}. Chiral plasma effects in astrophysics have also been explored in pulsars and black holes --  objects surrounded by magnetospheres made of plasma -- where the CME current,  $\mathbf{J}_{B}=\mu_{5}\mathbf{B}$, is supposed to exist, with repercussions on the propagation of helical modes \cite{Gorbar2}.

{Pulsars can provide a route for probing Lorentz violation (LV). {Indeed, pulsar timing was used to constrain coefficients of the Standard Model extension fermion sector \cite{Altschul} and of its gravitational extension \cite{Dong}}. These previous examples differ from the present investigation in the
sense that we now restrain the CPT-odd photon sector (CFJ term) by using pulsar timing data.} Lately, the influence of the axion on pulsar timing array (PTA) results has been investigated for a heavy axion model \cite{Moslem}, postinflationary axionlike particles \cite{Servant} {and detection of the stochastic gravitational-wave background (SGWB) by PTAs \cite{Khlopov}.} Using DMs, constraints on millicharged dark matter were derived using a dataset of radio pulsars \cite{Caputo}. {In a similar way to PTAs, constraints on the axion-photon coupling have also been examined in pulsar polarization arrays (PPA) aimed at measurements of the radio wave propagation in a time oscillating axion background \cite{LLR23}.} Significant variations in DM measurements can occur due to distinct factors, such as solar wind \cite{Tiburzi} and plasma turbulence in ISM \cite{Ferrière}.

In this work, we examine modifications of DM caused by  {the CFJ plasma electrodynamics}.  We establish constraints on the {magnitude on the CFJ factors {$K_{AF}^{0}$ and  $|\mathbf{K}_{AF}|$}} in the context of an ISM cold chiral plasma ruled by the relation (\ref{ChiralPlasma}). The group velocity and arrival time are rewritten in terms of the chiral {parameter}, and an effective DM takes place. Using pulsar timing datasets for distances, DMs, and RMs of five pulsars, namely, B1919+21, B1944+17, B1929+10,  B2016+28, and  B2020+28, the additional term is bounded, and the constraints for the five pulsars are examined and compared.

\textbf{\textit{Time delay and dispersion measure in MCFJ electrodynamics }} -- For a cold chiral plasma described by the purely timelike MCFJ electrodynamics, four refractive indices were obtained {in Ref. \cite{Filipe1}, where negative refraction was also discussed. In this work we will only consider {positive refractive indices, } }
{\begin{align}
	n_{R}  &  =-\frac{K_{AF}^{0}}{2\omega}+\sqrt{1-\frac{\omega_{p}^{2}}{\omega\left(\omega+\omega_{c}\right)}+\left(  \frac{K_{AF}^{0}}{2\omega}\right)  ^{2}},\\
	n_{L}  &  =\frac{K_{AF}^{0}}{2\omega}+\sqrt{1-\frac{\omega_{p}^{2}}{\omega\left(\omega-\omega_{c}\right)}+\left(  \frac{K_{AF}^{0}}{2\omega}\right)  ^{2}} , \label{n_timelike}
	\end{align}}
{associated with RCP and LCP waves}\footnote{Here, we considered $q=-e$, the electron charge present in cyclotron frequency $\omega_{c}$ in the refractive indices obtained in Ref. \cite{Filipe1}.}, respectively. The circularly polarized modes associated to the indices $n_{R}$ and $n_{L}$ can propagate at the group velocities, given by
{\begin{equation}
	v_{g} =\frac{2\omega\left(\omega\pm\omega_{c}\right)^{2}}{2\omega\left(\omega\pm\omega_{c}\right)^{2}\mp\omega_{c}\omega^{2}_{p}}\sqrt{1-\frac{\omega_{p}^{2}}{\omega\left(\omega\pm\omega_{c}\right)}+\left(
		\frac{K_{AF}^{0}}{2\omega}\right)  ^{2}}, \label{groupvel_CFJ}
	\end{equation}}
with the ($\pm$) related to RCP (LCP) waves, respectively. In the high frequency regime, $\omega \gg \omega_{p}$, one finds
\begin{equation}
\left(  v_{g}\right)  ^{-1}  \approx 1+\frac{\omega_{p}^{2}}%
{2\omega^{2}}+\frac{\left(K_{AF}^{0}\right)^{2}}{8\omega^{2}},\label{groupvel_CFJ2}
\end{equation}
with a chiral term in the power $\omega^{-2}$ contributing to the time delay (between one wave traveling in vacuum and the other in the chiral plasma) as
\begin{equation}
\tau=\frac{e^{2}}{2m\omega^{2}}\left(  \text{DM}%
+\text{DM}_{CFJ}^{(\bullet)}\right), \label{timedelay_CFJ}
\end{equation}
where we define a chiral effective dispersion measure,  
\begin{equation}
\text{DM}_{CFJ}^{(\bullet)}=\frac{m\left(K_{AF}^{0}\right)^{2}}{4e^{2}}d.\label{DMCFJ}%
\end{equation}
The frequency dependence is preserved in (\ref{timedelay_CFJ}), behaving as $\omega^{-2}$, as well as in the usual time delay (\ref{timedelay3}). 
	The effective dispersion measure (\ref{DMCFJ}) can be read as a correction due to the chiral parameter $K_{AF}^{0}$ in the usual dispersion measure $\text{DM}$, here ascribed to ordinary electrons. In this sense, observational DM deviations can be employed to estimate limits on chiral factor $K_{AF}^{0}$ magnitude.

Considering now the scenario of electromagnetic propagation cold plasma governed by the CFJ chiral vector (see Ref.~\cite{Filipe2}), the influence of the chiral vector in the time delay is investigated in the view of the cases in which it is parallel ($ \mathbf{K}_{AF} \parallel \mathbf{B}_{0}$) and orthogonal ($\mathbf{K}_{AF} \perp \mathbf{B}_{0}$) to the magnetic field.

For a chiral vector parallel to the magnetic field, the RCP and LCP modes are associated with the following refractive indices\footnote{As in the timelike case, we considered $q=-e$ for the electron charge.}:
\begin{align}
n_{L(R)}^{2} &= 1 - \frac{\omega_{p}^{2}}{\omega ( \omega \mp \omega_{c})} \pm \frac{\lvert \mathbf{K}_{AF}\rvert}{\omega}, \label{n_spacelike}
\end{align}
whose related group velocity (in the high-frequency regime),
\begin{equation}
	\left(  v_{g}\right)  ^{-1}  \approx 1+\frac{\omega_{p}^{2}}{2\omega^{2}}+\frac{\lvert \mathbf{K}_{AF}\rvert^{2}}{8\omega^{2}}.\label{groupvel_SL}
\end{equation}
has the same form as the group velocity (\ref{groupvel_CFJ2}), with  $\lvert \mathbf{K}_{AF} \rvert$ replacing $K_{AF}^{0}$. Thus, similarly, the time delay is
  \begin{equation}
  \tau=\frac{e^{2}}{2m\omega^{2}}\left(  \text{DM}%
  +\text{DM}_{CFJ}^{(\bullet\bullet)}\right), \label{timedelay_CFJ2}
  \end{equation}
with $\text{DM}_{CFJ}^{(\bullet\bullet)}$ defined as
  	\begin{equation}
  	\text{DM}_{CFJ}^{(\bullet\bullet)}=\frac{m\lvert \mathbf{K}_{AF} \rvert^{2}}{4e^{2}}d.\label{DMCFJ_SL}%
  	\end{equation}
  
As for the case the chiral vector is orthogonal to the magnetic field,  $ \mathbf{K}_{AF} \perp \mathbf{B}_{0}$, there are two refractive indices associated with elliptical polarizations \cite{Filipe2}, from which only one,
	\begin{equation}
	\left(n_{B}\right)^{2}=S-\frac{\lvert \mathbf{K}_{AF} \rvert^{2}}{2P\omega^{2}} +\frac{1}{P}\sqrt{P^{2}D^{2}+\frac{\lvert \mathbf{K}_{AF} \rvert^{4}}{4\omega^{4}}},\label{n_spacelike2}
	\end{equation}
produces relevant order contributions to the group velocity,
	\begin{equation}
	\left(  v_{g}\right)  ^{-1}  \approx 1+\frac{\omega_{p}^{2}}{2\omega^{2}}+ \frac{\lvert \mathbf{K}_{AF}\rvert^{2}}{2\omega^{2}}.
	\end{equation}   
In this configuration, the delay becomes
\begin{equation}
	\tau=\frac{e^{2}}{2 m\omega^{2}}\left(  \text{DM}%
	+4\text{DM}_{CFJ}^{(\bullet\bullet)}\right). \label{timedelay_CFJ3}
	\end{equation}
The effective $\text{DM}_{CFJ}$ above differs by a factor 4 from the one given in \eqref{DMCFJ_SL}, obtained in the configuration $\mathbf{K}_{AF} \parallel \mathbf{B}_{0}$, while the $\lvert \mathbf{K}_{AF}\rvert^{2}$ behavior is maintained.

\textbf{\textit{Dispersion measure constraints on the chiral parameters}} -- Equations (\ref{timedelay_CFJ}) and (\ref{timedelay_CFJ2}) represent the modified time delay for an ISM pervaded by an MCFJ plasma, which receives corrections in terms of the chiral dispersion measure $\text{DM}_{CFJ}$. The nature of such an additive contribution allows us to constrain the magnitude of the chiral factors using the observational data of pulsars. Indeed, by proposing that the observed DM is equal to the sum of the usual DM and the CFJ correction, $\text{DM}_{\text{obs}} = \text{DM}+\text{DM}_{CFJ}$, the chiral contribution can be limited by the observational uncertainties.

LOFAR census dataset provides observational dispersion measure values for several pulsars \cite{LOFAR}, from which, for our estimates, we have selected five, namely, B1919+21, B1944+17,  B1929+10, B2016+28, and B2020+28. The catalog also provides an uncertainty, denoted by $\epsilon_{\text{DM}}$, which, in our analysis, is ascribed to the chiral factor parameter. Thus, for each measurement and to restrain the chiral parameters ($K_{AF}^{0}$, $\mathbf{K}_{AF}$) magnitude, we impose
	\begin{equation}
\text{DM}_{CFJ}\lesssim\epsilon_{\text{DM}},
\label{constraint1}
	\end{equation} {assuming that the correction magnitude is limited by the uncertainty in DM measurements. As known, this is not the most rigorous approach, but it can be justified by the tiny size of the Lorentz-violating terms. In this sense, it allows establishing the first constraints on the chiral parameters using pulsar timing data.}
	
For the pulsars' distance from the Earth, $d$, appearing in (\ref{timedelay_CFJ}) and (\ref{timedelay_CFJ2}), we consider the \textit{corrected distances} listed in Ref.~\cite{Verbiest}, given in (k pc).

\textit{The timelike case} -- Using standard values for the constants, {the dispersion measure (\ref{DMCFJ}) provides, by condition (\ref{constraint1}), the following constraint:}
\begin{equation}
	K_{AF}^{0}  \lesssim \left(2.35\times 10^{-12}\text{eV}\right)\sqrt{\frac{\text{k pc} }{d}}\sqrt{\frac{\epsilon_{DM}}{\text{pc cm}^{-3}}}.\label{constraint_DM_timelike}
	\end{equation}
Starting with the pulsar B1919+21, the LOFAR census gives $\text{DM}_{\text{obs}}= 12.44399(63) \ \text{pc} \ \text{cm}^{-3}$, with $\text{pc} = 3.086\times10^{16}\text{m}$, and an error $\epsilon_{\text{DM}}=0.00063\ \text{pc} \ \text{cm}^{-3}$.

In addition, taking $d\approx0.3 \ \text{k pc}$,  the constraint (\ref{constraint_DM_timelike}) implies
\begin{equation}
	K_{AF}^{0} \lesssim  1.1\times 10^{-22} \ \text{GeV}. \label{restricao1}
	\end{equation}
Considering now the pulsar B1944+17, the catalog provides $\text{DM}_{\text{obs}}= 16.1356(73) \ \text{pc} \ \text{cm}^{-3}$, with $\epsilon=0.0073 \ \text{pc} \ \text{cm}^{-3}$, and $d\approx0.3 \ \text{k pc}$, providing
\begin{equation}
K_{AF}^{0} \lesssim   3.7\times 10^{-22} \ \text{GeV}.
	\end{equation}
Analogously, the other three pulsars, B1929+10, B2016+28, and B2020+28, appear with a different order of magnitude, resulting in a chiral factor limited to the order of $10^{-23}$ GeV. All data and constraints are presented in Table~\ref{tab:table_constraints}.

\begin{widetext}

	
	\begin{table}[htbb] 
		\centering
		\caption{Constraints on the chiral parameters using DM data.}
		\label{tab:table_constraints}
		
		    \renewcommand{\arraystretch}{1.2}
		\setlength{\tabcolsep}{12pt}
		
		\begin{tabular}{*{5}{c}}
			\toprule[0.8pt]\midrule
			\textbf{Pulsars} &  $\mathrm{DM}_{\mathrm{obs}}$ ($\mathrm{pc \,cm}^{-3}$)	& $d$ (k pc) & $K_{AF}^{0}$ and $\mathbf{K}_{AF}^{\parallel}$ (GeV) & $\mathbf{K}_{AF}^{\perp}$ (GeV)   \\ 
			\midrule
			B1919+21  &  $12.44399(63)$ & $0.3$ &  $1.1\times 10^{-22}$ & $5.3\times 10^{-23}$  \\ 
			B1944+17 & $16.1356(73)$   & $0.3$    &   $3.7\times 10^{-22}$ &  $1.8\times 10^{-22}$  \\ 
	
			B1929+10 &   $3.18321(16)$  & $0.31$ &    $5.3\times 10^{-23}$ &  $2.6\times 10^{-23}$  \\ 
		
			B2016+28 &  $14.1839(13)$  & $0.98$ &    $8.5\times 10^{-23}$ & $4.2\times 10^{-23}$   \\ 
		
			B2020+28  & $24.63109(18)$ & $2.1$ &   $2.2\times 10^{-23}$  & $1.9\times 10^{-23}$    \\
			\midrule\bottomrule[0.8pt]
		\end{tabular}

	\end{table}	
\end{widetext}

\textit{The spacelike case} -- For the case $\mathbf{K}_{AF}\parallel\mathbf{k}$, the estimated constraints on the scalar chiral factor $K_{AF}^{0}$ are also valid for the chiral vector $\mathbf{K}_{AF}$, since the chiral DM has the same form in for both cases, see Eqs. (\ref{DMCFJ}) and (\ref{DMCFJ_SL}), respectively. The results are shown in Table~\ref{tab:table_constraints}.

As for the case $\mathbf{K}_{AF}\perp\mathbf{k}$, the corresponding time delay is given in expression (\ref{timedelay_CFJ3}), which yields $4\text{DM}_{CFJ}^{(\bullet\bullet)}\lesssim\epsilon_{\text{DM}}$, implying 
\begin{align}
\label{restrction-V-vector-orthogonal-DM-1}
	\lvert\mathbf{K}_{AF}\rvert  \lesssim \left(1.174\times 10^{-12} \, \text{eV}\right)\sqrt{\frac{\text{k pc} }{d}}\sqrt{\frac{\epsilon_{DM}}{\text{pc cm}^{-3}}} , 
\end{align}
implying limitation of $\mathbf{K}_{AF}$  magnitude of the order $10^{-23}$~GeV, for the pulsars B1919+21, B1929+10, B2016+28, and B2020+28, and $|\mathbf{K}_{AF} | \lesssim 10^{-22}$ GeV  for the pulsar B1944+17. See the Table \ref{tab:table_constraints}.

\textbf{\textit{Rotation Measure constraints on the chiral parameters}} -- Let us focus on the refractive indices associated with circular polarizations, which occur in the timelike and spacelike  (with $\mathbf{K}_{AF}\parallel \mathbf{B}$) cases only. This constitutes an appropriate scenario to explore the Faraday rotation for waves propagating in ISM permeated by plasma chiral MCFJ plasma.

For the scalar chiral factor, taking the refractive indices given in \eqref{n_timelike} at second order in $1/\omega$, the associated wave vectors are
\begin{equation}
k_{R,L}\approx\omega{c}\mp \frac{K_{AF}^{0}}{2}-\frac{\omega_{p}^{2}}{2\omega}+\frac{\left(K_{AF}^{0}\right)^{2}}{8\omega}\pm\frac
{\omega_{c}\omega_{p}^{2}}{2\omega^{2}}.
\end{equation}
Using these wave vectors in \eqref{Faraday}), Faraday rotation assumes the form
\begin{equation}
	\Delta\phi  =\lambda^{2}\left(\text{RM}-\text{RM}^{\left(\bullet\right)}_{CFJ}\right),
\end{equation}
where we define the additional $\lambda$-dependent RM term
\begin{equation}
	\text{RM}^{\left(\bullet\right)}_{CFJ}=\frac{K_{AF}^{0}}{2\lambda^{2}}d,\label{RM_CFJ1}
\end{equation}
stemming from the presence of the chiral factor $K_{AF}^{0}$.

As for the  vector chiral factor, whose refractive indices are given in (\ref{n_spacelike}), the wave vectors for the configuration $\mathbf{K}_{AF}\parallel \mathbf{B}$ can be written as
	\begin{equation}
	k_{R,L}\approx\omega\mp \frac{\lvert \mathbf{K}_{AF}\rvert}{2}-\frac{\omega_{p}^{2}}{2\omega}-\frac{\lvert \mathbf{K}_{AF}\rvert^{2}}{8\omega}\pm\frac
	{\omega_{c}\omega_{p}^{2}}{2\omega^{2}}\mp \frac{\omega_{p}^{2}\lvert \mathbf{K}_{AF}\rvert}{4\omega^{2}}, 
	\end{equation}
which, in Eq. (\ref{Faraday}), yields
	\begin{equation}
	\Delta\phi  =\lambda^{2}\left(\text{RM}-\text{RM}^{\left(\bullet\bullet\right)}_{CFJ}\right),
	\end{equation}	
where the chiral vector defines an RM contribution 
	\begin{equation}
	\text{RM}^{\left(\bullet\bullet\right)}_{CFJ}=\frac{\lvert \mathbf{K}_{AF}\rvert}{2\lambda^{2}}d+\frac{e^{2}\lvert \mathbf{K}_{AF}\rvert}{4\kappa }\text{DM},\label{RM_CFJ2}
	\end{equation}
with $\kappa=4\pi^{2}m$. The first term in the right-handed side of \eqref{RM_CFJ2} is similar to the expression (\ref*{RM_CFJ1}) with $\lvert \mathbf{K}_{AF}\rvert$ in the place of $K_{AF}^{0}$, while the second one involves the DM associated. 

\textit{Constraints using RM} -- Performing the same procedure of the last section for DMs data, we consider the measurement uncertainties given in the LOFAR data for RMs \cite{LOFAR_RM}, here denoted by $\epsilon_{\text{RM}}$, as the upper magnitude for the chiral contribution in (\ref{RM_CFJ1}) and (\ref{RM_CFJ2}), that is, $\text{RM}_{CFJ}\lesssim\epsilon_{\text{RM}}$. The respective constraints in both scalar and vector chiral factors' magnitude read 
\begin{equation}
	K_{AF}^{0}  \lesssim \left(5.19\times 10^{-26}\text{eV}\right)\left(\frac{\text{k pc}}{d}\right)\left(\frac{\epsilon_{\text{RM}}}{\text{rad m}^{-2
	}}\right),
	\end{equation}
and
\begin{equation}
	| \mathbf{K}_{AF} |  \lesssim \frac{\left(5.19\times 10^{-26}\text{eV}\right)\left(\epsilon_{\text{RM}}/\text{rad m}^{-2
		}\right)}{\left(d/\text{k pc}\right) + 1.9\times 10^{-12}\left(\text{DM}/\text{pc cm}^{-3}\right)}.
	\end{equation}

In the latter, the wavelength was associated with the \textit{centre frequency} of 148.9 MHz ($\lambda\approx 2.01338$ m), at which the pulsar observations were recorded in Ref.~\cite{LOFAR_RM}. {For the same five pulsars} B1919+21, B1944+17, B1929+10, B2016+28, and B2020+28, we obtain equal constraints on the chiral parameters $K_{AF}^{0}$ and $\mathbf{K}_{AF}^{\parallel}$, both at the order of $10^{-36}$ -- $10^{-37}$ GeV, as presented in the Table \ref{tab:table_constraints2}.


\begin{table}[H] 
	\centering
	\caption{Constraints on the chiral parameters using RM data.}
	\label{tab:table_constraints2}
	\begin{tabular}{*{4}{c}} 
	\toprule[0.8pt]\midrule
		\textbf{Pulsars} &  $\mathrm{RM}_{obs}$ {($\mathrm{rad~ m}^{-2}$)} 	& $d$ (k pc) & $K_{AF}^{0}$ and $\mathbf{K}_{AF}^{\parallel}$ (GeV)    \\ 
		\midrule
		B1919+21  &  $-15.04 \pm 0.02$ & $0.3$ &  $3.4\times 10^{-36}$ \\ 		B1944+17 & $-43.64 \pm 0.02$   & $0.3$    &   $3.4\times 10^{-36}$ \\ 
		B1929+10 &   $-5.27 \pm 0.01$  & $0.31$ &    $1.6\times 10^{-36}$ \\		B2016+28 &  $-33.14 \pm 0.01$  & $0.98$ &    $5.3\times 10^{-37}$   \\ 
		B2020+28  & $-72.56\pm 0.02$ & $2.1$ &   $5.0\times 10^{-37}$  \\
	\midrule\bottomrule[0.8pt]	
	\end{tabular}
	
\end{table}

\textbf{\textit{{Final remarks \label{conclusion}}}} -- A chiral cosmic medium, addressed as an ISM plasma ruled by the MCFJ electrodynamics \cite{Filipe1, Filipe2},  was investigated in aspects concerned with dispersion and rotation measure in order to impose astrophysical limits on the { magnitude of the LV chiral parameters}, by using pulsar timing of five pulsars: B1919+21, B1944+17, B1929+10,  B2016+28, and  B2020+28. For such a medium, a modified time delay (\ref{timedelay_CFJ}) with a chiral dispersion measure contribution, $\text{DM}_{CFJ}$, was obtained. The chiral parameters $K_{AF}^{0}$ and $\lvert \mathbf{K}_{AF}\rvert$ (for parallel configuration) are bounded according to \eqref{constraint_DM_timelike}, while $\lvert \mathbf{K}_{AF}\rvert$ (for orthogonal configuration) is restricted by \eqref{restrction-V-vector-orthogonal-DM-1}. Using dispersion measure (DM) data from the selected pulsars, the $K_{AF}^{0}$ and ${|\bf{K}}_{AF}^{\parallel}|$ magnitudes were restrained to the order $10^{-22}$ GeV, while $|\mathbf{K}_{AF}^{\perp}|\lesssim 10^{-23}$ GeV. Further, using rotation measure (RM) from the same pulsars, we have stated $K_{AF}^{0}$, ${|\bf{K}}_{AF}^{\parallel}|$  $\lesssim 10^{-36}$ GeV (see Table \ref{tab:table_constraints2}).

While the constraints on $K_{AF}^{0}$ (or $K_{AF}^{0}$) and $\lvert \mathbf{K}_{AF}\rvert$ (or $|{\bf{K}}_{AF}^{\parallel}|$), using DM data, are not as restrictive as the ones {from CMB polarization \cite{Caloni,DataTables_Kostelecky} and astrophysical birefringence~\cite{CFJ}}, {they are competitive in comparison with the bounds envolving Schumann resonances \cite{Mewes-bounds-1}, resonant cavities \cite{Yuri} and solar wind data \cite{Spallicci}. Moreover, they constitute the} first ones on sideral chiral plasma parameters using dispersion measure of pulsars (located in our galaxy at distances of {a few parsecs}).  On the other hand, the RM data (for the same pulsars) provided tighter bounds, improved by 14 orders of magnitude in relation to the {DM ones} (see Table.~\ref{tab:table_constraints2}). {Table~\ref{tab:table_constraints3} compares our constraints from DM and RM data with the scenarios mentioned above. The RM contraints improve by 12 orders of magnitude the bounds stemming from solar wind investigation, the most restrictive one apart from CBM polarization and {astrophysical birefringence,} representing competitive restrictions. }
 
{As a final point, we comment on the possible connections with the axion dark matter.} In the cold dark matter scenario, only the time dependence for the axion field is considered, with $\theta=\theta_{0}e^{im_{a}t}$ {(the axion oscillates on a timescale of 1/mass), with} $\theta_{0}=g_{a\gamma\gamma}\sqrt{\rho_{a}}/m_{a}$, where $\rho_{a}$ and $m_{a}$ are the local axion dark matter density and mass, respectively. In addition,  $g_{a\gamma\gamma}$ is the axion-photon coupling constant \cite{CAPP,Cast,Dwarf}, which can be estimated with our constraints on the chiral parameter $K_{AF}^{0}$. {At the same time, the cold axion space variation is neglected ($\nabla \theta =\mathbf{0}$) over the distance scale of the system ($L$), which is fulfilled requiring that the Compton length is larger than ($L$), that is, $\lambda>L$. In this case, the axion field does not depend on the space coordinates. {As is well-known}, MCFJ electrodynamics becomes equivalent to the axion theory when the space derivative $\partial_{\mu}\theta=(K_{AF})_{\mu}$ is considered constant. For the zero component, $\partial_{t}\theta=K_{AF}^{0}$, this assumption} is realized when timescales are
	much shorter than the period of the axion oscillation, where $m_{a}t\ll 1$, yielding $\theta\approx\theta_{0}m_{a}t$ \cite{Peter}. In doing so, we have $\partial_{t}\theta\approx\theta_{0}m_{a}$, and the coupling constant can be written as $g_{a\gamma\gamma}= K_{AF}^{0}/\sqrt{\rho_{a}}$. {It is important to note that the pulsars used in Table~\ref{tab:table_constraints} and Table~\ref{tab:table_constraints2} are located at distances smaller than 1 kpc and the axion space derivative should be constant over such a distance. This implies that the axion mass should be lighter than $10^{-27}$ eV, which is much smaller than the axion mass range favored for dark matter (whose mass should be larger than $\sim 10^{-22}$ eV) \cite{HOTW17}. In this sense, it is not possible to suppose that the constraints of Table~\ref{tab:table_constraints} and Table~\ref{tab:table_constraints2} hold for cold axion dark matter. }
	
\begin{table}[t] 
	\centering
	\caption{{Bounds comparison among distinct setups.}}
	\label{tab:table_constraints3}
	\begin{tabular}{*{4}{c}} 
		\toprule[0.8pt]\midrule
		\textbf{---} &  $K_{AF}^{0}$ (GeV)	&  $\mathbf{K}_{AF}$ (GeV) & Ref.   \\ 
		\midrule
		{CMB polarization} & $10^{-45}$ & - &   \cite{Caloni} \\
		{Astrophys. Birefringence} & $10^{-42}$  & $10^{-42}$ &   \cite{CFJ} \\
		\textbf{Pulsar RM}  & $10^{-36}$ & $10^{-36}$ &   \textbf{Table \ref{tab:table_constraints2}} \\
		Solar wind &  -  & $10^{-24}$ &   \cite{Spallicci} \\ 
		\textbf{Pulsar DM}  &  $10^{-22}$ &  $10^{-23}$ &  \textbf{Table \ref{tab:table_constraints}}\\
		Resonant cavities &   -- &  $10^{-23}$ &    \cite{Yuri} \\	
		Schumann resonances & $10^{-21}$   & $10^{-21}$   &   \cite{Mewes-bounds-1} \\ 
		\midrule\bottomrule[0.8pt]	
	\end{tabular}
\end{table}

In summary, we have shown that pulsar timing data may be useful in further investigations concerning chiral plasmas { and their optical properties.}


\textit{\textbf{Acknowledgments}}. The authors thank FAPEMA, CNPq, and CAPES (Brazilian research agencies) for their invaluable financial support. M.M.F. is supported by FAPEMA APP-12151/22, CNPq/Produtividade 317048/2023-6 and CNPq/Universal/422527/2021-1. P.D.S.S. is grateful to FAPEMA APP-12151/22. Furthermore, we are indebted to CAPES/Finance Code 001 and FAPEMA/POS-GRAD-04755/24.

\end{document}